\documentclass[12pt]{iopart}

\pdfoutput=1
\usepackage{iopams}
\usepackage{graphicx}
\usepackage[colorlinks=true, urlcolor=blue, linkcolor=blue, citecolor=blue, pdftex]{hyperref}
\usepackage{subfig}
\usepackage{float}

\begin{document}

\title[Valence-bond crystals in the kagom\'e spin-$1/2$ Heisenberg antiferromagnet ...]{Valence-bond crystals in the kagom\'e spin-$1/2$ Heisenberg antiferromagnet: a symmetry classification and projected wave function study}

\author{Yasir Iqbal$^1$, Federico Becca$^2$ and Didier Poilblanc$^1$}
\address{${}^{1}$ Laboratoire de Physique Th\'eorique UMR-5152, CNRS and Universit\'e de Toulouse, F-31062 Toulouse, France}
\address{${}^{2}$  Democritos National Simulation Center, Istituto Officina dei Materiali del CNR, Via Bonomea 265, I-34136 Trieste, Italy}

\eads{\mailto{iqbal@irsamc.ups-tlse.fr}, \mailto{becca@sissa.it},\mailto{didier@irsamc.ups-tlse.fr}}

\begin{abstract}
In this paper, we do a complete classification of 
valence-bond crystals (VBCs) on the kagom\'e lattice based on general arguments 
of symmetry only and thus identify many new VBCs for different unit cell
sizes. For the spin-$1/2$ Heisenberg antiferromagnet, we study the relative 
energetics of competing gapless spin liquids (SLs) and VBC phases within the class 
of Gutzwiller-projected fermionic wave functions using variational Monte Carlo
techniques, hence implementing exactly the constraint of one fermion per site.
By using a state-of-the-art optimization method, we conclusively show that the
U($1$) Dirac SL is remarkably stable towards dimerizing into 
all $6$-, $12$- and $36$-site unit cell VBCs. This stability is also 
preserved on addition of a next-nearest-neighbor super-exchange coupling of 
both antiferromagnetic and ferromagnetic (FM) type. However, we find that a 
$36$-site unit cell VBC is stabilized on addition of a very small 
next-nearest-neighbor FM super-exchange coupling, i.e. 
$|J_{2}|\approx0.045$, and this VBC is the same in terms of space-group 
symmetry as that obtained in an effective quantum dimer model study. 
It breaks reflection symmetry, has a nontrivial flux pattern and is a 
strong dimerization of the uniform RVB SL. 
\end{abstract}


\submitto{\NJP}

\maketitle

\section{Introduction}

For many decades, physicists have been actively searching for playgrounds 
that are `hot' enough to melt magnetic freezing at temperatures well below 
the characteristic interaction energy scales in the system. This melting being
fueled by quantum fluctuations leads to stabilization of exotic quantum 
paramagnetic phases of matter~\cite{Balents-2010}. Representatives of such 
phases are spin liquids (SLs) and valence-bond crystals (VBCs); the former 
preserve lattice symmetries and the latter break them, according to a generally
accepted definition. Long before any experimental hints, theoreticians such 
as Pomeranchuk already conjectured the existence of SLs~\cite{Pomeranchuk-1941},
which were later advocated by Anderson to be possible appropriate ground states
for the spin-$1/2$ Heisenberg antiferromagnet~\cite{Anderson-1973,Anderson-1987}. 
On the experimental side, the drought in the search for SLs ended with the 
discovery of Herbertsmithite (${\rm Zn}{\rm Cu}_{3}({\rm OH})_{6}{\rm Cl}_{2}$),
a compound with perfect kagom\'e lattice geometry, belonging to the 
paratacamite family~\cite{Shores-2005,Bert-2007,Lee-2007,Lee-2008,deVries-2008,Imai-2008,Olariu-2008,Mendels-2010,Han-2011}. 
In it, the combination of low spin value ($S=1/2$), low-dimensionality ($d=2$)
and coordination number ($z=4$) and frustrating nearest-neighbor (NN) 
antiferromagnetic (AF) super-exchange interactions on a non-bipartite lattice 
leads to amplification of quantum fluctuations that stabilize a 
quantum paramagnet. 
Indeed, all experimental probes on Herbertsmithite point to a SL behavior
down to $20$ mK ($\sim J/10^{4}$), which was established on the magnesium 
version of Herbertsmithite (i.e. ${\rm Mg}{\rm Cu}_{3}({\rm OH})_{6}{\rm Cl}_{2}$)~\cite{Mendels-2007,Helton-2007,Kermarrec-2011}.
Furthermore, Raman spectroscopic studies on Herbertsmithite hint at
a gapless (algebraic) SL~\cite{Wulferding-2010}. 

On the theoretical side however, the nature of the ground state of the NN 
spin-$1/2$ quantum Heisenberg antiferromagnet (QHAF) on the kagom\'e lattice is 
still elusive and intensely debated. Exact diagonalization studies 
have revealed a magnetically disordered ground state and a huge number of 
singlet excitations below the triplet gap~\cite{Elser-1989,Zeng-1990,Chalker-1992,Leung-1993,Elstner-1994,Lecheminant-1997,Waldtmann-1998,Sindzingre-2000,Waldtmann-2000,Richter-2004,Sorensen-2009,Sindzingre-2009,Nakano-2011,Lauchli-2011}. 
Using approximate numerical techniques various claims as to the nature of the 
ground state have been made. These have included, among SL phases, a gapless 
(algebraic) U($1$) Dirac SL using projected fermionic variational Monte 
Carlo~\cite{Ran-2007,Hermele-2008,Ma-2008,Iqbal-2010,Iqbal-2011}, a gapped 
${\mathbb Z}_{2}$ SL~\cite{Sachdev-1992,Wang-2006,Lu-2011} using density matrix 
renormalization group (DMRG)~\cite{Jiang-2008,Yan-2011} and a chiral 
topological SL using Schwinger boson mean field theory~\cite{Messio-2011}. 
Among the VBC phases, the proposals have included a $36$-site unit cell 
VBC~\cite{Marston-1991,Nikolic-2003} numerically studied using series 
expansion~\cite{Singh-2007,Singh-2008}, multi-scale entanglement 
renormalization ansatz (MERA)~\cite{Evenbly-2010} and also using quantum dimer
models (QDM)~\cite{Zeng-1995,Poilblanc-2010,Schwandt-2010}. Furthermore, 
VBCs with smaller unit cells of $6$ sites~\cite{Budnik-2004}, 
$12$ sites~\cite{Hastings-2000,Syro-2002,Syro-2004} and $18$ 
sites~\cite{Marston-1991} were also argued to be viable ground states of the 
spin-$1/2$ QHAF. A more recent generalized QDM study found a new (possibly 
chiral) VBC of $12$-site unit cell to be competing with the $36$-site unit cell VBC. 
It also established an extensive quasi-degeneracy of the ground state 
manifold of the kagom\'e $S=1/2$ QHAF with a stiff competition between several 
phases~\cite{Poilblanc-2011}.

In this work, we will study these non-magnetic phases within a Schwinger 
fermion formulation of the spin model. Within this approach, the 
{\it projected} gapless (algebraic) U($1$) Dirac SL has the best 
variational energy~\cite{Ran-2007}; despite being a marginally stable 
phase, it was argued in~\cite{Hermele-2008} to be stable against a certain 
class of perturbations. Explicit numerical calculations using projected wave 
functions have in fact shown it to be stable ({\it locally} and {\it globally})
w.r.t. dimerizing into all known VBC perturbations~\cite{Ran-2007,Ma-2008,Iqbal-2010}. 
Furthermore, it was shown that within this class of Gutzwiller projected wave 
functions, all the fully symmetric gapped ${\mathbb Z_{2}}$ SLs have a 
higher energy compared to the U($1$) Dirac SL~\cite{Iqbal-2011}. Similar 
conclusions were also reached within the Schwinger boson approach to the spin 
model~\cite{Tay-2011,Yang-2012}. Note that a simple tensor network (PEPS) representation of such a projected bosonic RVB ansatz can be constructed and has been studied in~\cite{Poilblanc-2012}.

In this paper, in section~\ref{sec:symmetry} we first perform a systematic 
symmetry classification of VBC patterns on the kagom\'e lattice and thus identify
and enumerate many new VBCs, independent of the formalism used to study them. 
In section~\ref{sec:numerics}, we address the question
of relative energetics of SL and VBC phases. In particular, in 
section~\ref{sec:Dirac-stability} we show that the U($1$) Dirac SL is 
remarkably stable w.r.t. dimerizing into any of these new VBCs. This 
stability is also preserved upon addition of a finite next-nearest-neighbor
(NNN) super-exchange coupling of both AF and ferromagnetic (FM) type. 
Such a NNN coupling might be a possible perturbation in Herbertsmithite. 
In section~\ref{sec:RVB-stability}, we show that a broken symmetry phase is 
stabilized on addition of a small NNN FM coupling, which is consistent with the
findings in~\cite{Ralko-2010}. This VBC has a $36$-site unit cell with a 
non-trivial flux pattern threading its plaquettes and it is found to be a 
strong dimerization of another competing U($1$) gapless SL, the so-called
uniform RVB SL~\cite{Ran-2007}. This $36$-site unit cell VBC has 
a lower symmetry as compared to that studied in our previous 
work~\cite{Iqbal-2010} and has precisely the same symmetry as that 
identified in QDM studies~\cite{Poilblanc-2010,Schwandt-2010,Poilblanc-2011}. 
Thus, here we mainly establish 
the stability of the U($1$) Dirac SL w.r.t. an extremely large class of 
potential VBC instabilities and detect a non-trivial $36$-site unit cell VBC instability of 
the uniform RVB SL which is stabilized on addition of a very weak NNN 
FM super-exchange coupling to the Hamiltonian.

\subsection{The model, wave functions and the numerical technique}
The Hamiltonian for the spin-$1/2$ quantum Heisenberg $J_{1}{-}J_{2}$ model is
\begin{equation}
\label{eqn:heis-ham}
\hat{{\cal H}} = J_{1} \sum_{\langle ij \rangle} \mathbf{\hat{S}}_{i} \cdot \mathbf{\hat{S}}_{j} + J_{2} \sum_{{\boldsymbol \langle}\langle ij \rangle{\boldsymbol \rangle}} \mathbf{\hat{S}}_{i} \cdot \mathbf{\hat{S}}_{j}
\end{equation} 
where $\langle ij \rangle$ and 
${\boldsymbol \langle}\langle ij \rangle{\boldsymbol \rangle}$ denote sums
over NN and NNN pairs of sites, respectively. The $\mathbf{\hat{S}}_{i}$ are 
spin-$1/2$ operators at each site $i$. In the following, we will consider 
$J_{1}>0$ (AF) and both FM and AF super-exchange $J_{2}$; all energies will be 
given in units of $J_{1}$.

The physical variational wave functions are defined by projecting 
noncorrelated fermionic states:
\begin{equation}
\label{eqn:var-wf}
|\Psi_{{\rm VMC}}(\chi_{ij},\Delta_{ij},\mu,\zeta)\rangle=\mathbf{{\cal P}_{G}}|\Psi_{{\rm MF}}(\chi_{ij},\Delta_{ij},\mu,\zeta)\rangle,
\end{equation}
where $\mathbf{{\cal P}_{G}}=\prod_{i}(1-n_{i,\uparrow}n_{i,\downarrow})$ is
the full Gutzwiller projector enforcing the one fermion per site constraint.
Here, $|\Psi_{{\rm MF}}(\chi_{ij},\Delta_{ij},\mu,\zeta)\rangle$ is the ground
state of a mean-field Hamiltonian constructed out of Schwinger fermions and 
containing hopping, chemical potential and singlet pairing terms:
\begin{equation}
\label{eqn:MF0}
{\cal H}_{{\rm MF}} = \sum_{i,j,\alpha} (\chi_{ij}+\mu\delta_{ij})c_{i,\alpha}^{\dagger}c_{j,\alpha}
+\sum_{i,j} \{(\Delta_{ij}+\zeta\delta_{ij})c^{\dagger}_{i,\uparrow}c^{\dagger}_{j,\downarrow}+h.c.\} \, ,
\end{equation}
where $\chi_{ij}=\chi_{ji}^{*}$ and $\Delta_{ij}=\Delta_{ji}$. Besides the 
chemical potential $\mu$, we will also consider real and imaginary components 
of on-site pairing, which are absorbed in $\zeta$.

\begin{figure}
\centering
\subfloat[U($1$) Dirac spin liquid Ansatz]{\label{fig:DSL-ansatz}\includegraphics[width=0.47\columnwidth]{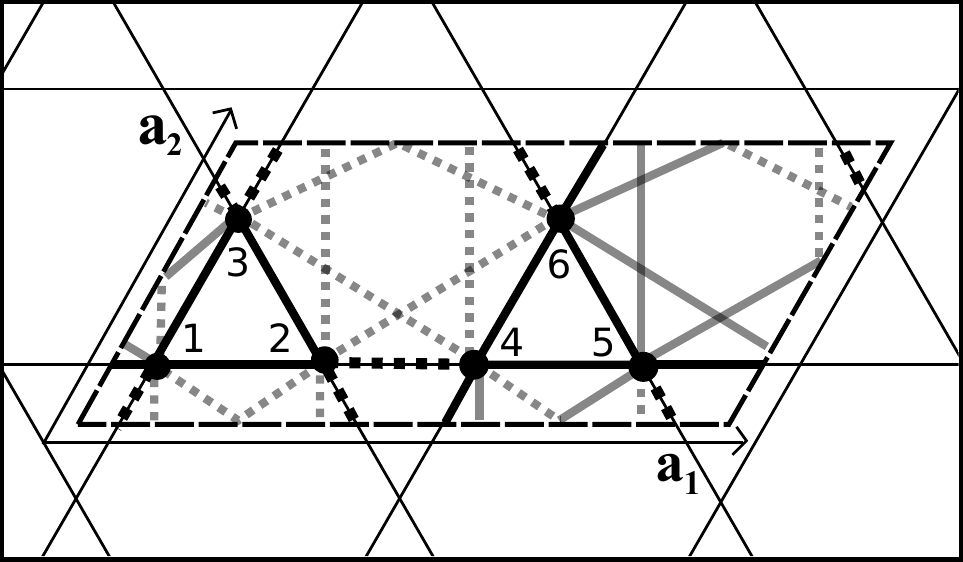}}\hspace{0.5cm}                
\subfloat[CVBC]{\label{fig:CVBC-6}\includegraphics[width=0.47\columnwidth]{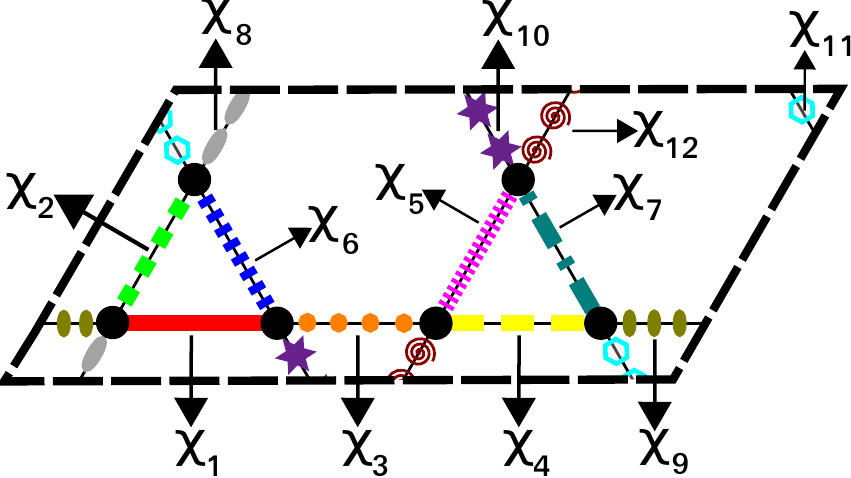}}
\caption{(a) The U($1$) Dirac SL ansatz given up to second NN bonds. The unit 
cell has to be doubled to accommodate the $\pi$-flux. The black (gray) bonds 
denote first NN real hopping (second NN real hopping) terms. The solid (dashed)
bonds denote positive (negative) hoppings. (b) The columnar VBC has no 
(point group) symmetries at all, hence all its $12$ bonds are different, which are thus 
marked with different colors and line styles. Consequently, its symmetry (point) group
is the identity {\it E}.}
\label{fig:DSL-CVBC}
\end{figure}

\begin{figure}
\centering
\includegraphics[width=1.0\columnwidth]{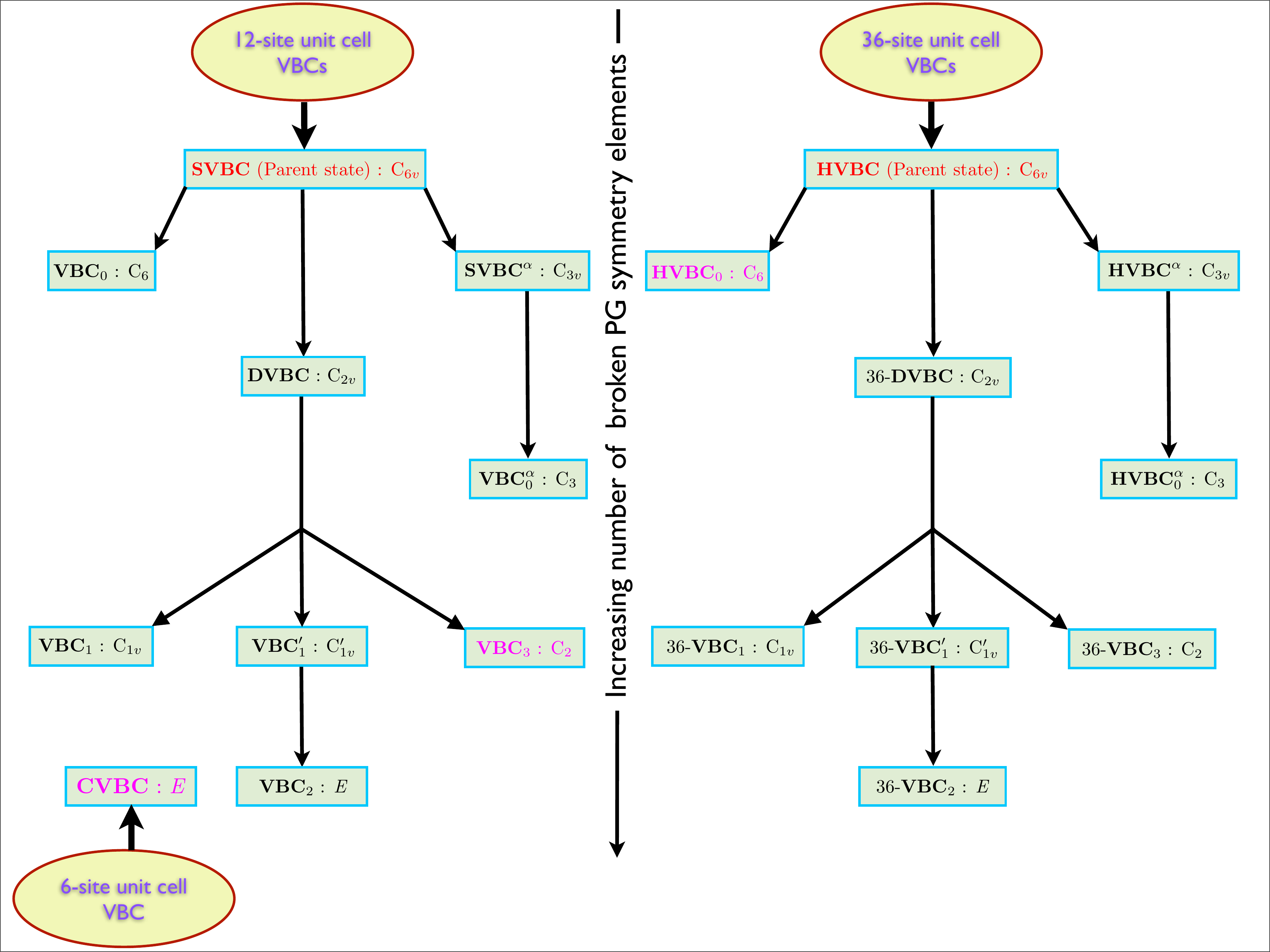} 
\caption{A hierarchical flowchart sorting out the myriad of different $6$-, 
$12$- and $36$-site unit cell VBCs in order of increasing (from top to bottom)
number of broken point group (PG) symmetry elements. The square boxes contain 
the VBC names followed by their respective symmetry PG. The `parent'
(maximally symmetric) VBCs are marked in red and those which have been found
as competing ground states in studies using quantum dimer models are marked 
in pink~\protect\cite{Poilblanc-2010,Schwandt-2010,Poilblanc-2011}. The corresponding 
VBC patterns, and their discussion, are given in the text. As much as possible, we use 
labeling for the VBCs which is similar to that used in~\protect\cite{Poilblanc-2011}.}
\label{fig:vbc-chart}
\end{figure}

The SL phases are characterized by different patterns of distribution of 
underlying SU($2$) gauge fluxes through the plaquettes which are implemented 
by a certain distribution of the phases of $\chi_{ij}$ and $\Delta_{ij}$ on the
lattice links. Since in a SL state $|\chi_{ij}|^2+|\Delta_{ij}|^2$ is constant
for each geometrical distance, a complete specification of a SL state up to 
$n$th NN amounts to specifying, in addition to the SU($2$) fluxes, the optimized 
magnitude of hopping and pairing parameters at each geometrical distance and 
the specification of the on-site terms $\mu$ and $\zeta$~\cite{Wen-1991,Wen-2002}.
On the other hand, in a VBC state $|\chi_{ij}|^2+|\Delta_{ij}|^2$ may be
different from bond to bond and, therefore, the specification of VBCs amounts 
to giving the pattern of amplitudes of $\chi_{ij}$ and $\Delta_{ij}$ at each 
geometrical distance, in addition to specifying the SU($2$) fluxes through the
plaquettes.These parameters are the {\it ans\"atze} of a given state and serve
as the variational parameters in the physical wave function that are optimized
within the variational Monte Carlo scheme to find the energetically best state.
It is worth mentioning that we use a sophisticated implementation of the 
stochastic reconfiguration (SR) optimization method which allows us to obtain 
an extremely accurate determination of variational 
parameters~\cite{Sorella-2005,Yunoki-2006}. Indeed, small energy differences 
are effectively computed by using a correlated sampling, which makes it 
possible to strongly reduce statistical fluctuations. This feature is 
especially important for the spin-$1/2$ QHAF since the energies of all the 
competing phases are rather close.

\subsection{Parent spin liquid states}
The ansatz for the energetically best variational state, the U($1$) Dirac SL, 
is given in figure~\ref{fig:DSL-ansatz}. Due to the U($1$) flux $\varphi$ 
being $0$ and $\pi$ [$\exp{(i\varphi)}=\prod_{{\rm plaquette}}\chi_{ij}$] 
through triangles and hexagons, respectively, it is denoted as $[0,\pi]$ SL. 
In its mean-field band structure the Fermi surface collapses to two points 
at which the spectrum becomes relativistic with Dirac conical 
excitations~\cite{Ran-2007}. Another energetically competing state, the uniform
RVB SL, has zero flux through all plaquettes and is therefore denoted as $[0,0]$
SL. Its mean-field band structure consists of large circular spinon Fermi 
surfaces~\cite{Ma-2008}. Both these states are fully symmetric, U($1$) gapless
SLs and can be extended to include second NN hoppings, leading to a gain in energy 
without changing their nature~\cite{Iqbal-2010}. It is worth 
noting that the effect of projection on these mean-field states can be 
drastic. 

\section{Symmetry classification and enumeration of valence-bond crystals (VBCs)}
\label{sec:symmetry}
The VBC states on the kagom\'e lattice break its elementary (three-site) unit cell
translation symmetry with different unit cell sizes, which describe their 
modulation. In previous studies~\cite{Budnik-2004,Hastings-2000,Syro-2002,Syro-2004,Marston-1991,Nikolic-2003,Singh-2007,Evenbly-2010,Poilblanc-2010,Schwandt-2010,Poilblanc-2011}, 
using different methods, VBCs with $6$-, $12$-, $18$- and $36$-site unit cells 
were identified as possible ground states of the spin-$1/2$ QHAF. In this work,
we will restrict our analysis to VBCs with $6$-, $12$- and $36$-site unit cells.
For each unit cell size with a given center of symmetry, we enumerate VBCs 
starting from the maximally symmetric (${\rm C}_{6v}$) `parent' VBC and 
systematically break point group symmetry elements, right down to the VBC with
no symmetry at all. This results in an enumeration of $19$ VBCs in total, 
$9$ VBCs each for $12$- and $36$-site unit cells and $1$ VBC for the $6$-site 
unit cell (see figure~\ref{fig:vbc-chart}). Only $6$ out of the $19$ VBC have 
been studied previously. In this paper, we will study the possibility of any of 
these VBCs to occur as the ground state.

\begin{figure}
\centering
\subfloat[SVBC]{\label{fig:SVBC-12}\includegraphics[width=0.3\columnwidth]{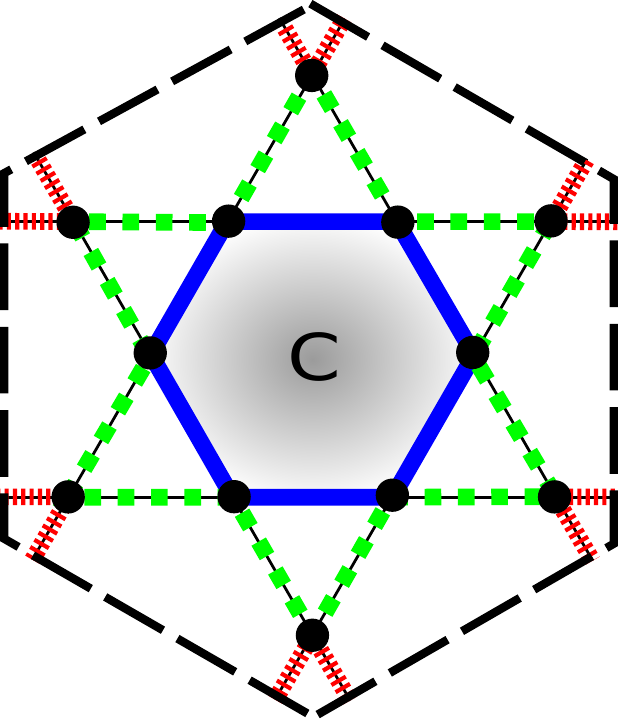}}\hspace{0.5cm}                
\subfloat[${\rm VBC}_{0}$]{\label{fig:VBC_0-12}\includegraphics[width=0.3\columnwidth]{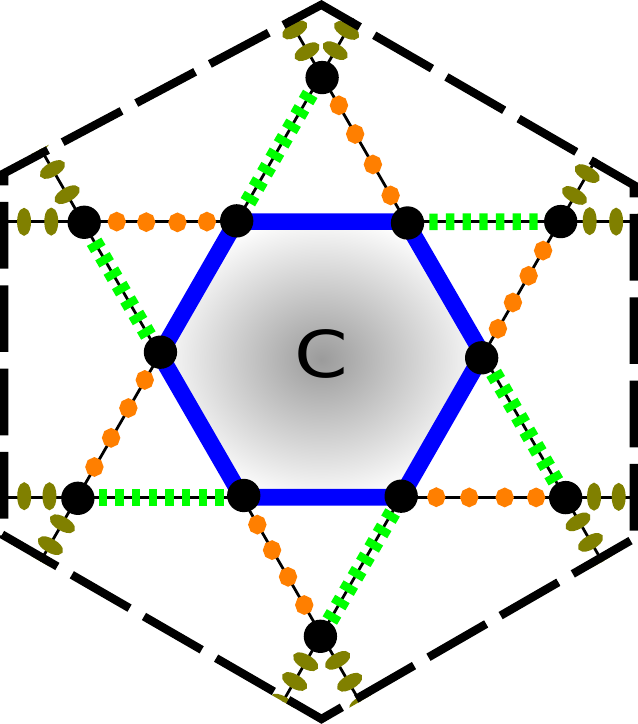}}\hspace{0.5cm}
\subfloat[${\rm SVBC}^{\alpha}$]{\label{fig:SVBC-alpha-12}\includegraphics[width=0.3\columnwidth]{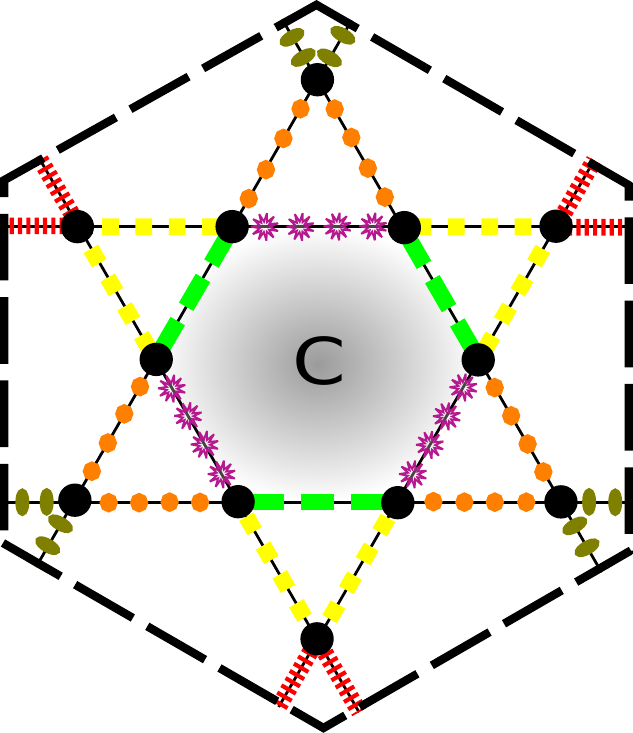}}
\caption{The most symmetric $12$-site unit cell VBCs: the center of symmetry is marked `C' 
(the center of the shaded hexagon), around which bonds connected by the given
PG symmetry operations are marked with the same color and style of the
line. We will henceforth refer to these bonds as being in the same {\it class}.
(a) The Star-VBC has the maximal PG symmetry, ${\rm C}_{6v}$; hence 
it acts as a `parent' VBC. Its bonds breakup into three distinct classes. 
(b) The ${\rm VBC}_{0}$ lacks crystallographic axes reflection symmetries 
in contrast to the SVBC; thus its symmetry group is reduced to ${\rm C}_{6}$. 
It has four classes of bonds. (c) The Star-${\rm VBC}^{\alpha}$ has 
reduced ($2\pi/3$) rotation symmetry but preserves reflection symmetry; thus 
its symmetry group is ${\rm C}_{3v}$. It has six classes of bonds.}
\label{fig:12-site-VBC-first-set}
\end{figure}

\begin{figure}[t]
\centering
\subfloat[${\rm VBC}_{0}^{\alpha}$]{\label{fig:VBC_0-alpha-12}\includegraphics[width=0.3\columnwidth]{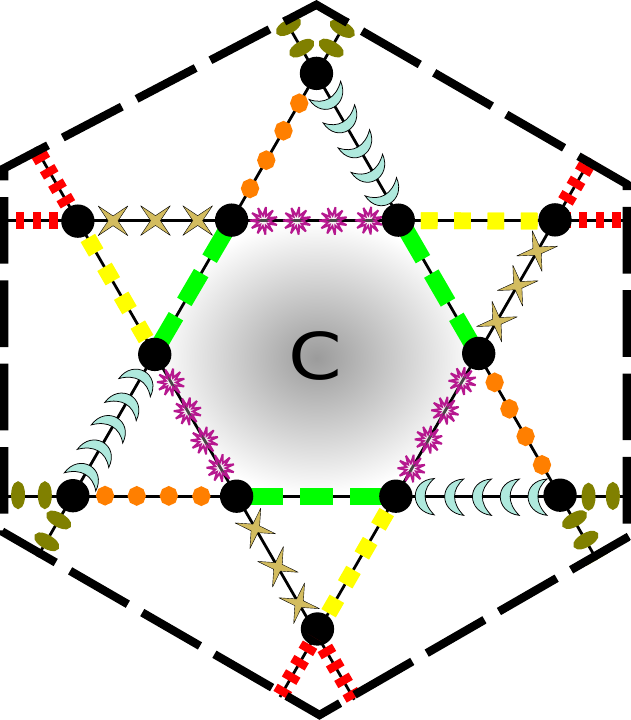}}\hspace{0.5cm}
\subfloat[DVBC]{\label{fig:DVBC-12}\includegraphics[width=0.3\columnwidth]{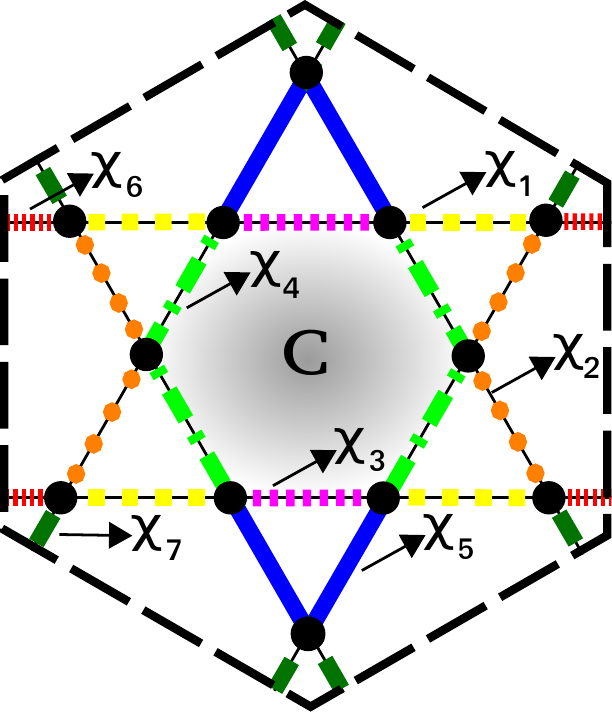}}\hspace{0.5cm}
\subfloat[${\rm VBC}_{3}$]{\label{fig:VBC_3-12}\includegraphics[width=0.3\columnwidth]{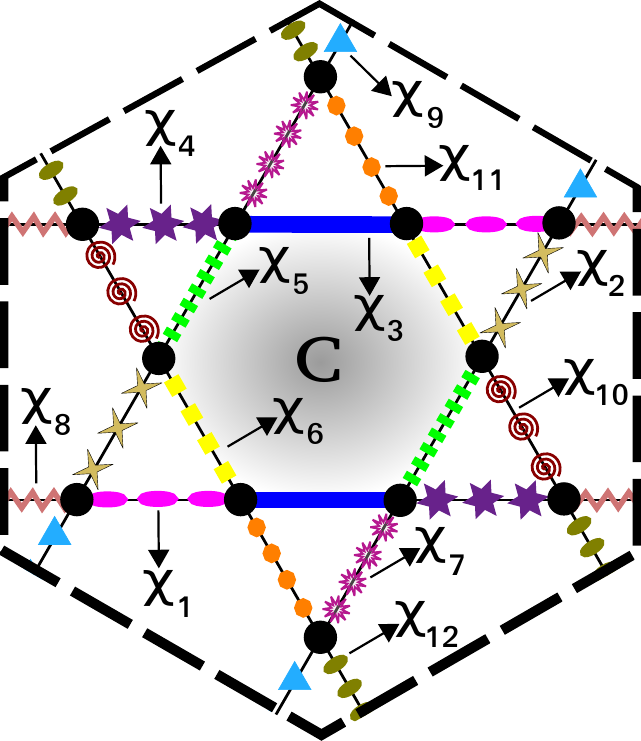}}\vspace{-0.15cm}
\subfloat[${\rm VBC}_{1}$]{\label{fig:VBC_1-12}\includegraphics[width=0.3\columnwidth]{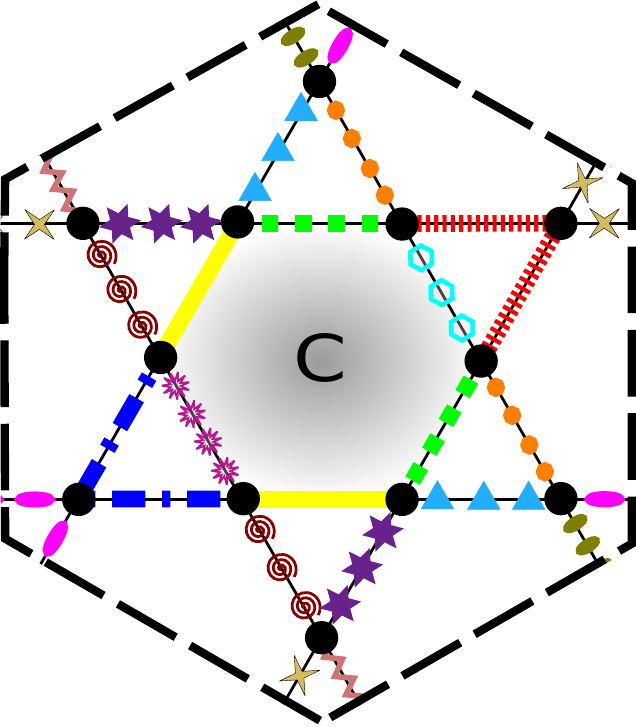}}\hspace{0.5cm}
\subfloat[${\rm VBC}_{1}'$]{\label{fig:VBC_1-prime-12}\includegraphics[width=0.3\columnwidth]{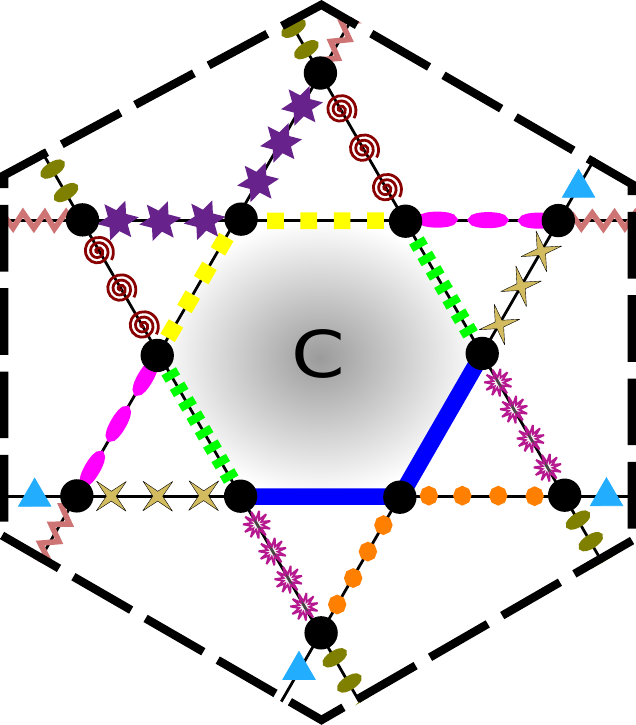}}\hspace{0.5cm}
\subfloat[${\rm VBC}_{2}$]{\label{fig:VBC_2-12}\includegraphics[width=0.3\columnwidth]{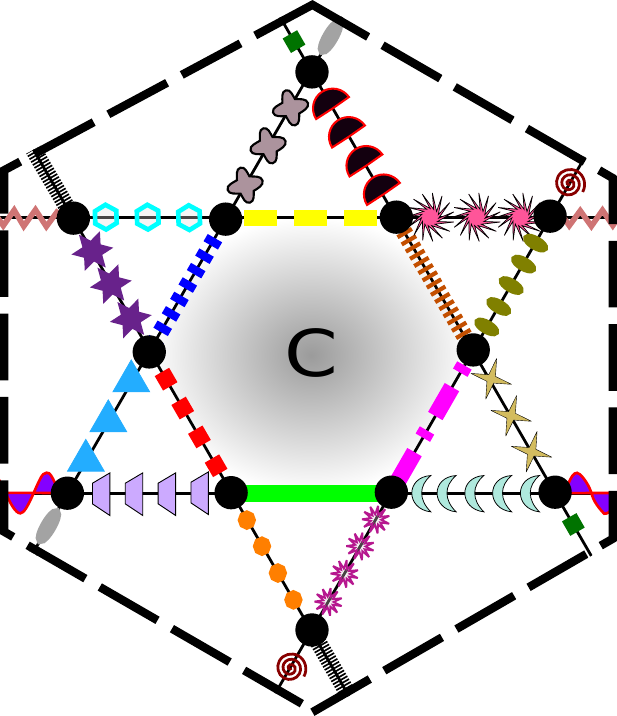}}
\caption{Other $12$-site unit cell VBCs: (a) the ${\rm VBC}_{0}^{\alpha}$ has only reduced rotation symmetry 
($2\pi/3$); thus in contrast to ${\rm SVBC}^{\alpha}$ its symmetry group is 
reduced to ${\rm C}_{3}$. It has eight classes of bonds. (b) The diamond-VBC has
two perpendicular axes of reflection symmetry, thus giving rise to 
${\rm C}_{2v}$ symmetry, with seven classes of bonds. (c) The ${\rm VBC}_{3}$ 
has only $\pi$-rotation symmetry; thus its symmetry group is ${\rm C}_{2}$. 
It has $12$ classes of bonds. (d) The ${\rm VBC}_{1}$ possesses only a single 
axis of reflection symmetry which bisects the sides of the shaded hexagon; 
consequently, its symmetry group is ${\rm C}_{1v}$. It has $14$ classes of 
bonds. (e) The ${\rm VBC}_{1}'$ has the same symmetry as ${\rm VBC}_{1}$, but 
its reflection symmetry axis passes through a vertex of the shaded hexagon; 
we shall denote the symmetry group as ${\rm C}_{1v}'$ to distinguish it from 
that of ${\rm VBC}_{1}$. It has $12$ classes of bonds. (f) The ${\rm VBC}_{2}$
has no symmetry whatsoever; hence its symmetry group is just the identity, 
denoted here as {\it E}. Consequently, it has $24$ distinct classes of bonds.}
\label{fig:12-site-VBC-second-set}
\end{figure}

\subsection{$12$-site unit cell VBCs}
The kagom\'e lattice can be viewed as a triangular lattice of $12$-site blocks 
shaped in the form of `stars'. Within this picture, it was argued 
in~\cite{Syro-2002,Syro-2004} that the ground state of the spin-$1/2$ QHAF has possible long-range
singlet order that settles in this triangular star arrangement. 
This lends support to the picture that the ground state can be a VBC with a $12$-site 
unit cell capturing some modulation. In total, nine symmetry distinct VBCs with 
a $12$-site unit cell can occur; see figures~\ref{fig:12-site-VBC-first-set} 
and~\ref{fig:12-site-VBC-second-set} for their NN patterns. In particular, 
the SVBC state (figure~\ref{fig:SVBC-12}) was argued in~\cite{Hastings-2000} 
to occur as an instability of the U($1$) Dirac SL and to be consequently 
stabilized as the ground state of the NN spin-$1/2$ QHAF. Numerical studies using 
projected wave functions have shown this proposal to be incorrect and have 
also established the stability of the uniform RVB SL w.r.t. dimerizing into 
the SVBC state~\cite{Ran-2007,Ma-2008,Iqbal-2010}. Furthermore, a recent QDM 
study~\cite{Poilblanc-2011} found the ${\rm VBC}_{3}$ state (figure~\ref{fig:VBC_3-12}) to be a 
competing ground state and a DMRG study~\cite{Yan-2011} concluded that the DVBC 
state (figure~\ref{fig:DVBC-12}) is close by in a generalized parameter space. In section~\ref{sec:numerics},
we study the possibility of a ground state realization of ${\rm VBC}_{3}$ and 
DVBC states numerically, within the framework of projected wave functions. 
In fact, we perform this study for {\it all} $12$-site unit cell VBCs.

\begin{figure}[t!]
\centering
\subfloat[HVBC]{\label{fig:HVBC-36}\includegraphics[width=0.3\columnwidth]{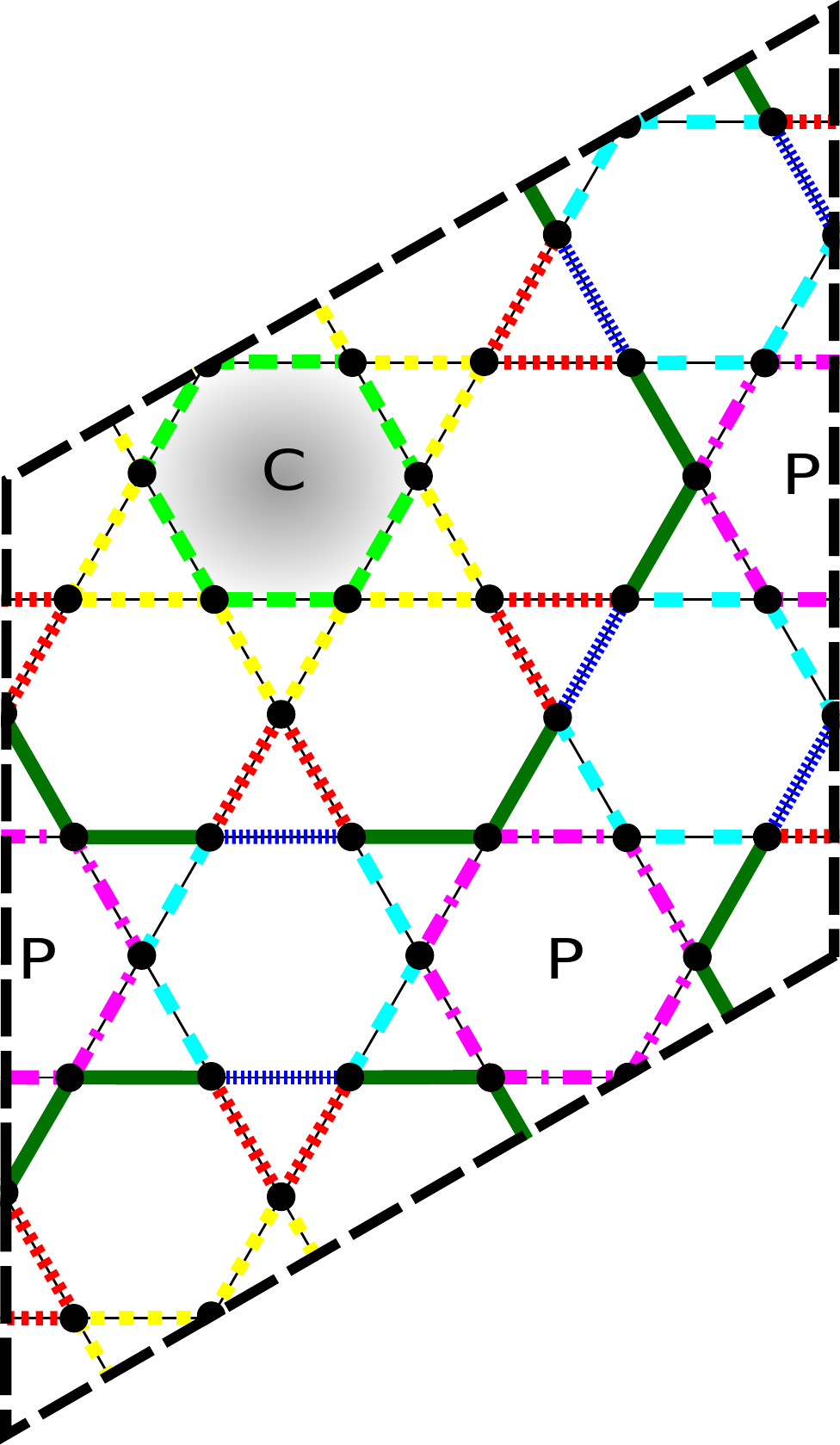}}\hspace{0.5cm}                
\subfloat[${\rm HVBC}_{0}$]{\label{fig:HVBC_0-36}\includegraphics[width=0.3\columnwidth]{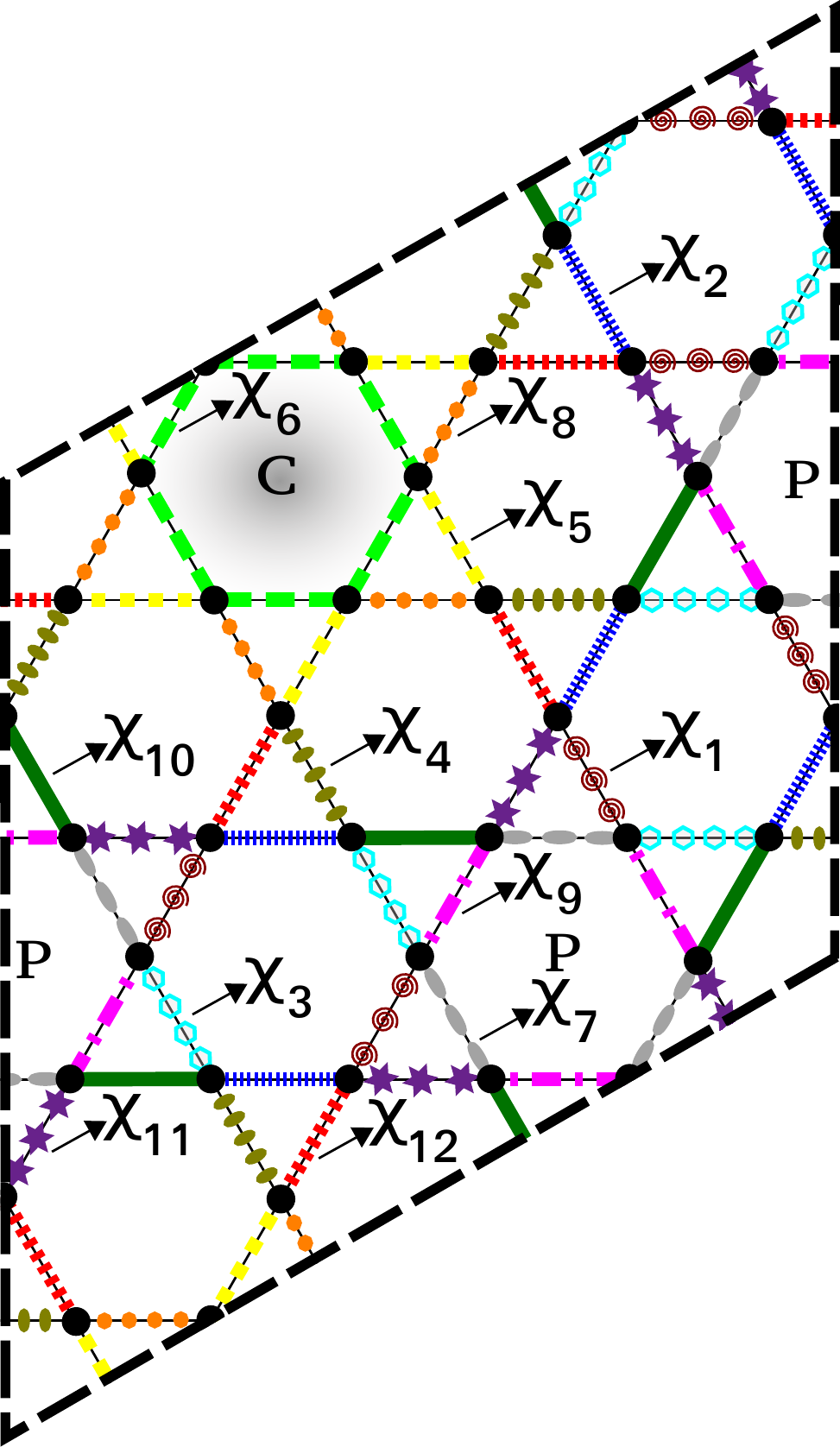}}\hspace{0.5cm}
\subfloat[${\rm HVBC}^{\alpha}$]{\label{fig:HVBC-alpha-36}\includegraphics[width=0.3\columnwidth]{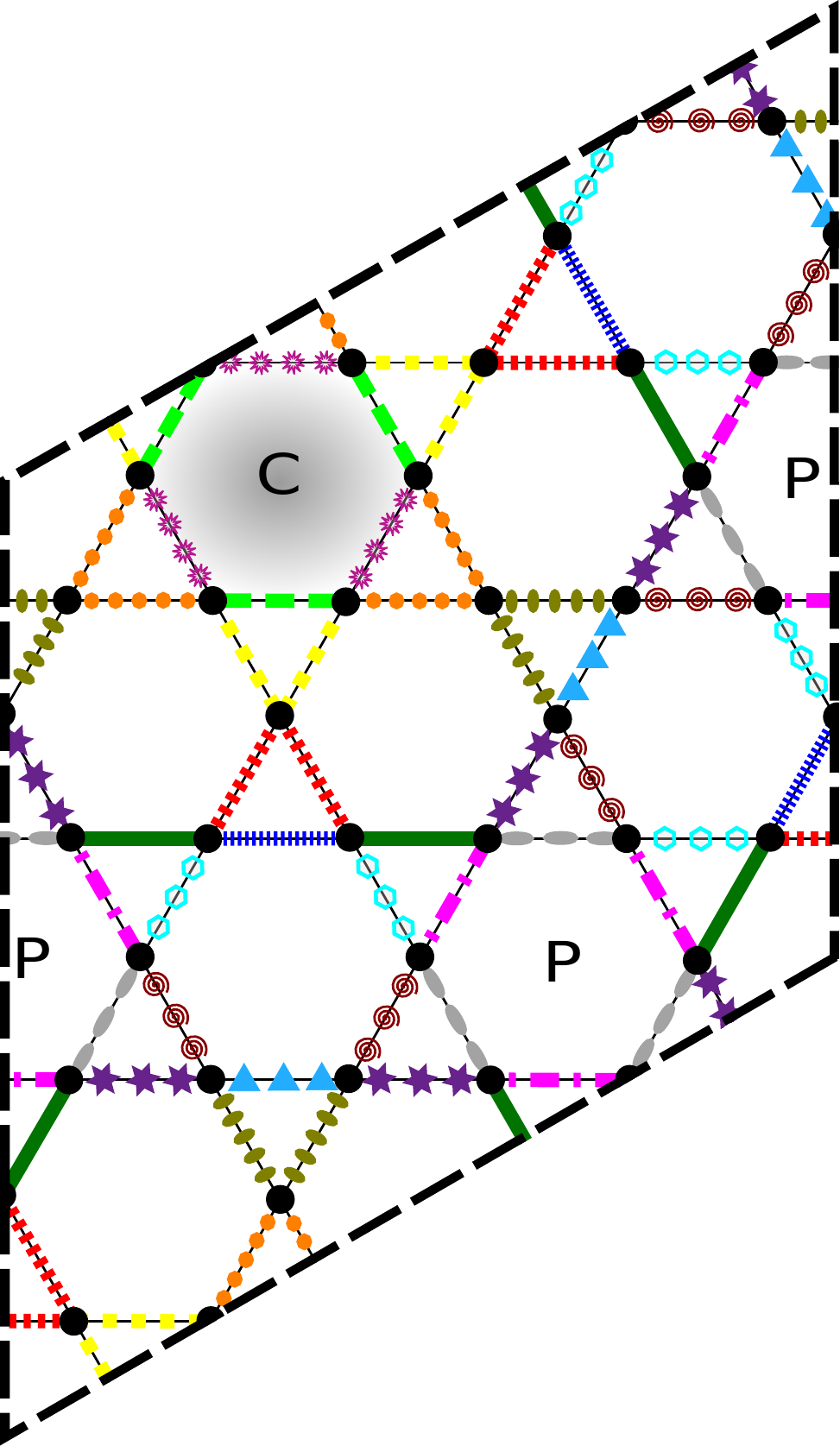}}\hspace{0.5cm}
\caption{The $36$-site unit cell VBCs: the center of symmetry is marked `C' 
(the center of the shaded hexagon). The {\it perfect} hexagons, marked at their centers by `P', 
form a honeycomb lattice at the center of which lie the shaded hexagons.\protect\footnotemark 
(a) The hexagonal-VBC has the maximal PG symmetry, ${\rm C}_{6v}$; 
hence it acts as a `parent' VBC. Its bonds breakup into seven distinct classes.
(b) The hexagonal-${\rm VBC}_{0}$, in contrast to the HVBC, lacks reflection 
symmetries about crystallographic axes; thus its symmetry group is reduced to 
${\rm C}_{6}$. It has $12$ classes of bonds. (c) The ${\rm HVBC}^{\alpha}$ 
has reduced ($2\pi/3$) rotation symmetry but preserves reflection symmetry; 
thus its symmetry group is ${\rm C}_{3v}$. It has $14$ classes of bonds.}
\label{fig:36-site-VBC-first-set}
\end{figure}
\footnotetext{The points P are {\it not} centers of inversion ($\pi$-rotation) symmetry, as has been mismarked in figure~1(a) of~\protect\cite{Poilblanc-2011}, which corresponds to the ${\rm HVBC}_{0}$ state in the present work.}

\begin{figure}
\centering
\subfloat[${\rm HVBC}_{0}^{\alpha}$]{\label{fig:HVBC_0-alpha-36}\includegraphics[width=0.3\columnwidth]{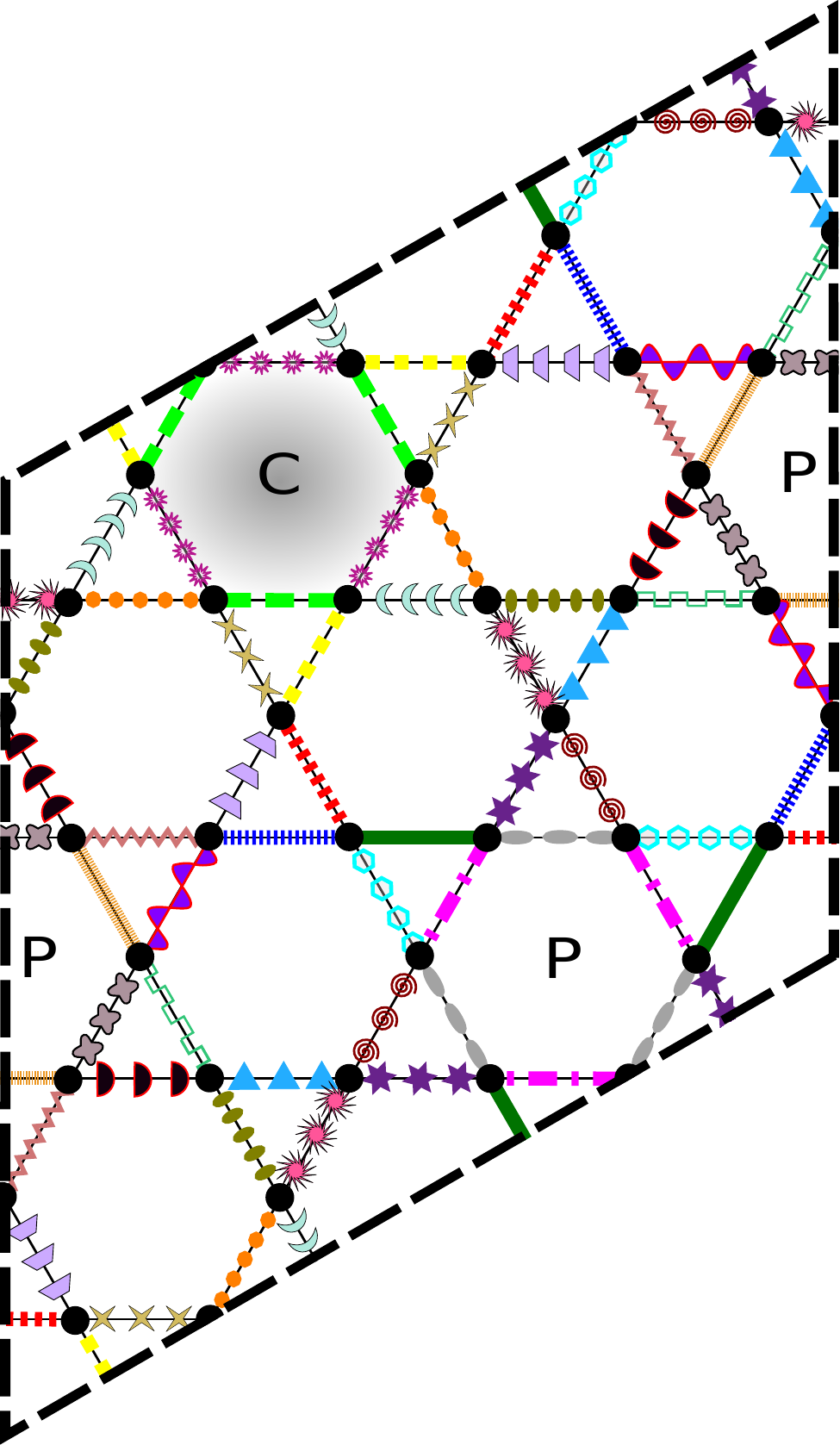}}\hspace{0.5cm}
\subfloat[$36$-DVBC]{\label{fig:DVBC-36}\includegraphics[width=0.3\columnwidth]{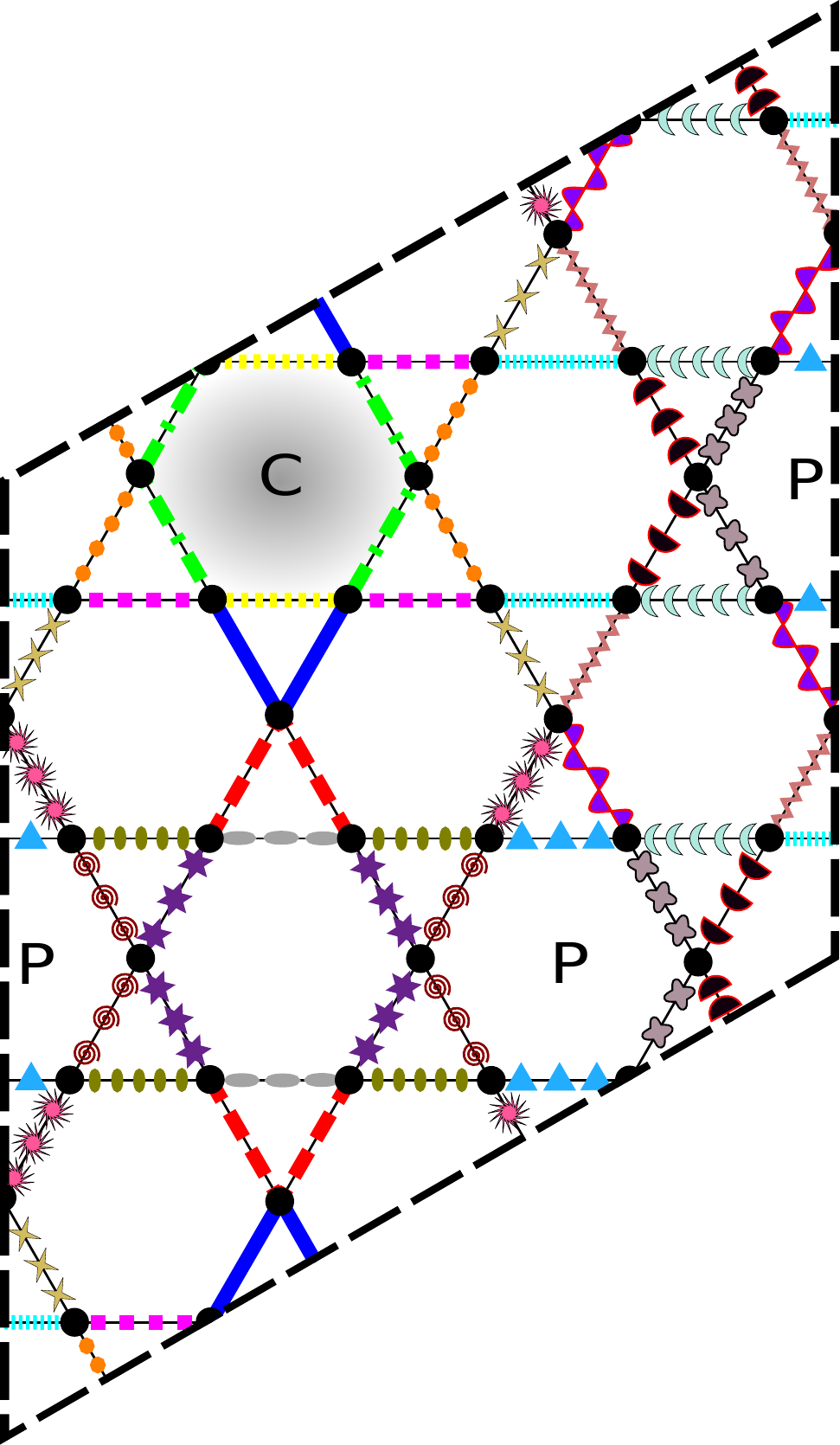}}\hspace{0.5cm}
\subfloat[$36$-${\rm VBC}_{3}$]{\label{fig:VBC_3-36}\includegraphics[width=0.3\columnwidth]{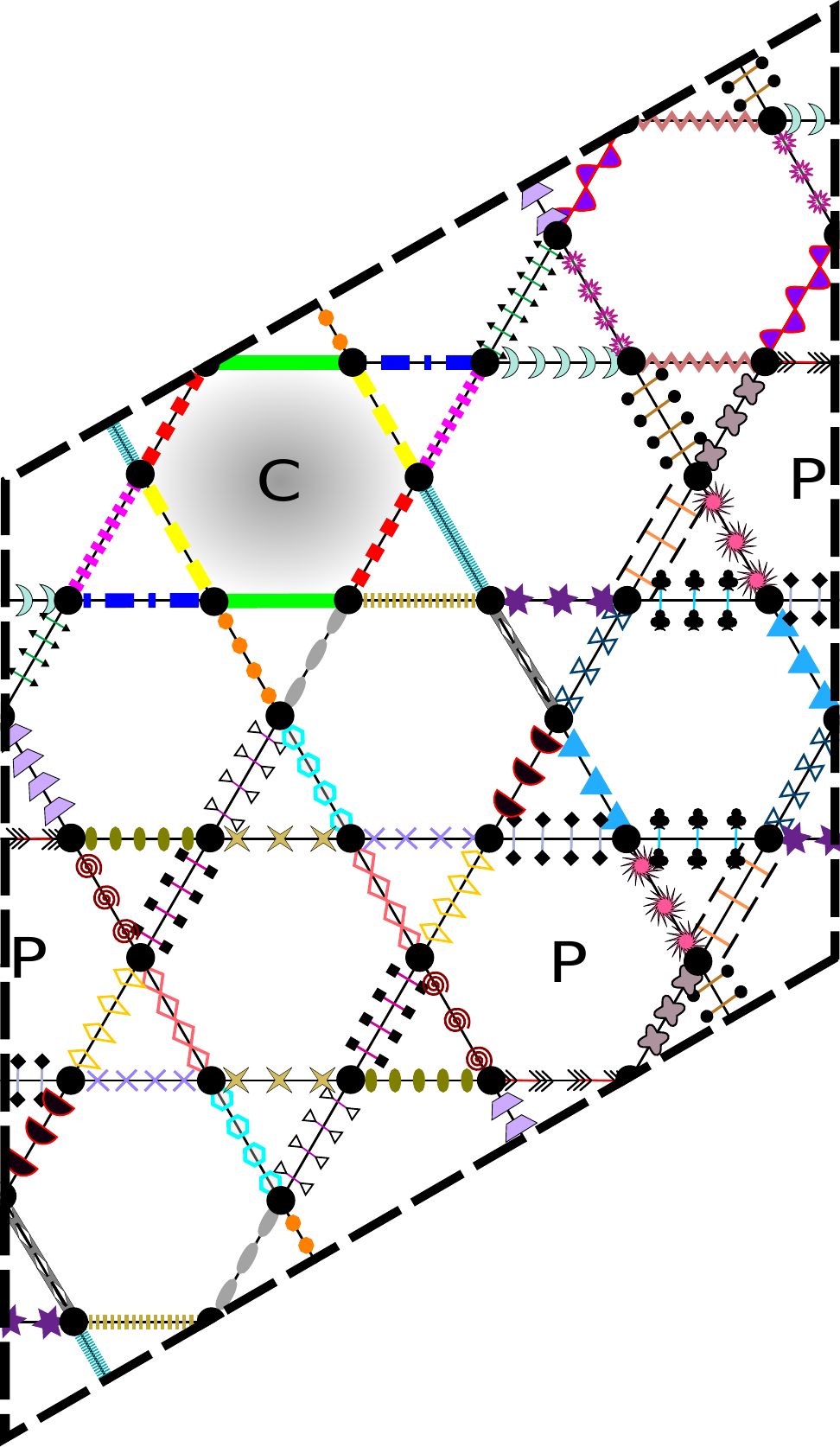}}
\caption{(a) The ${\rm HVBC}_{0}^{\alpha}$ has only a reduced rotation symmetry 
($2\pi/3$); thus in contrast to ${\rm HVBC}^{\alpha}$ its symmetry group is 
reduced to ${\rm C}_{3}$. It has $24$ classes of bonds. 
(b) The $36$-diamond-VBC has two perpendicular axes of reflection symmetry, 
thus giving rise to ${\rm C}_{2v}$ symmetry, with $19$ classes of bonds. 
(c) The $36$-${\rm VBC}_{3}$ has only $\pi$ rotation symmetry; thus its 
symmetry group is ${\rm C}_{2}$. It has $36$ classes of bonds.}
\label{fig:36-site-VBC-second-set}
\end{figure}

\begin{figure}
\centering
\subfloat[$36$-${\rm VBC}_{1}$]{\label{fig:VBC_1-36}\includegraphics[width=0.3\columnwidth]{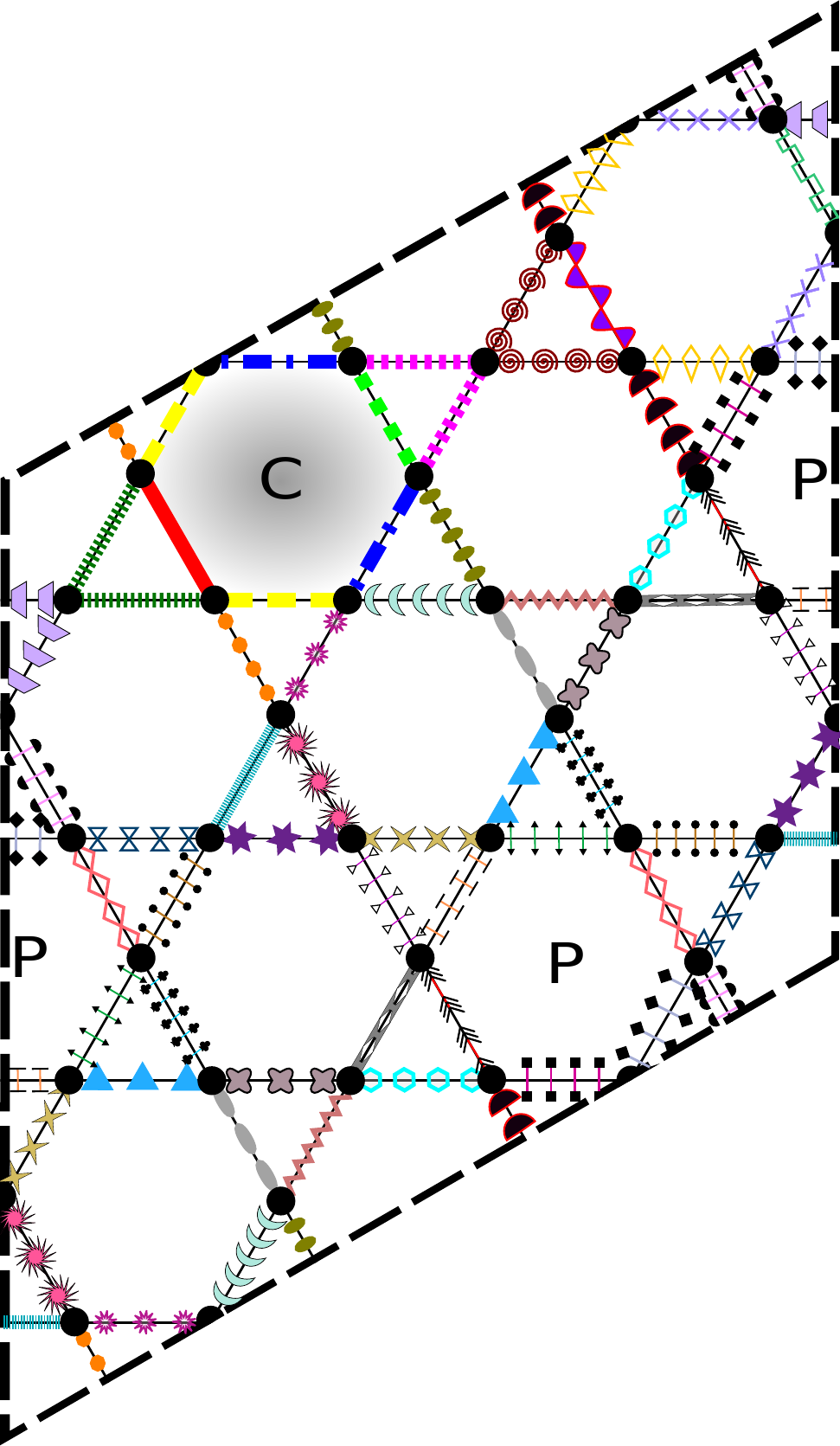}}\hspace{0.5cm}
\subfloat[$36$-${\rm VBC}_{1}'$]{\label{fig:VBC_1-prime-36}\includegraphics[width=0.3\columnwidth]{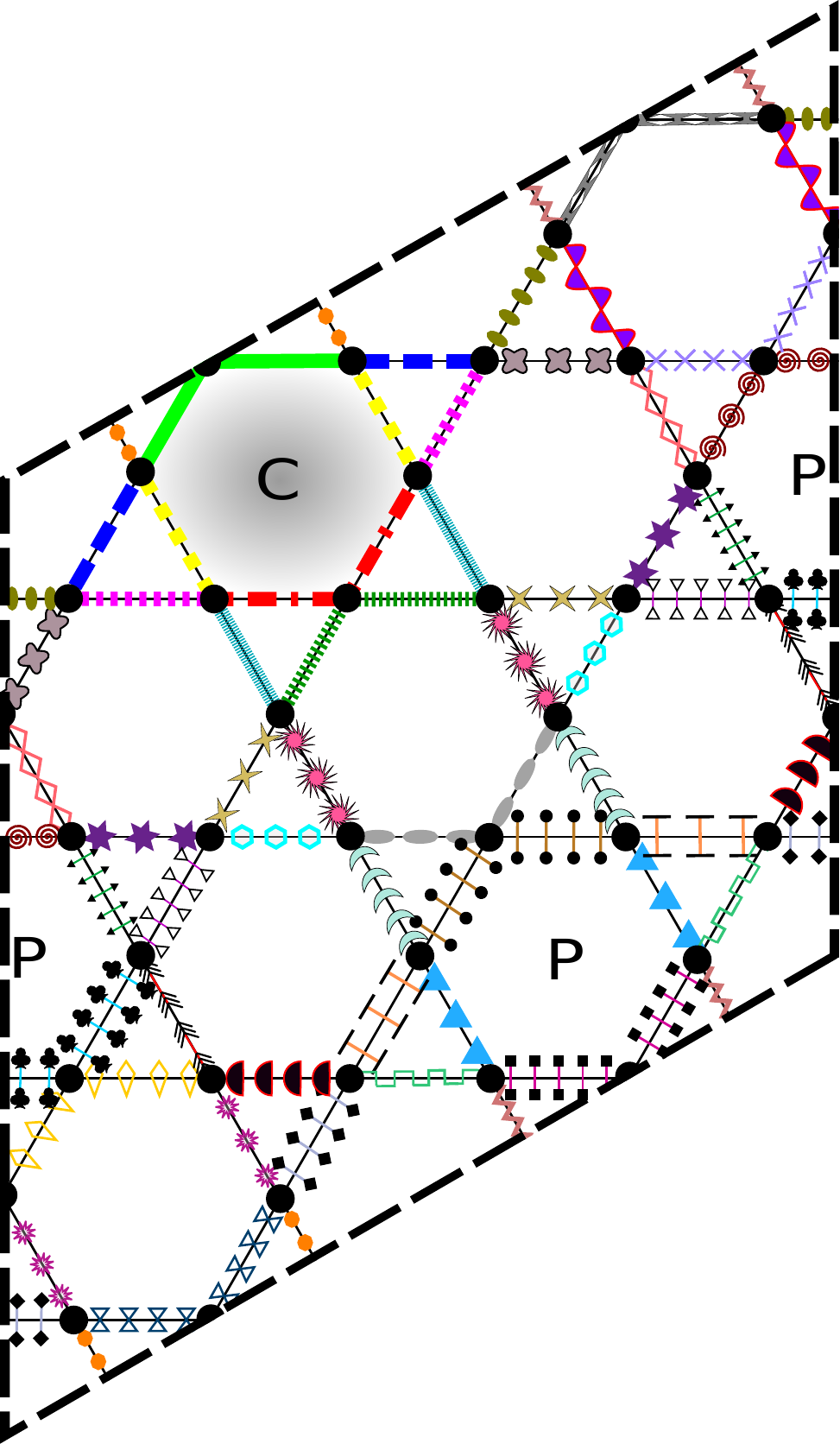}}
\caption{(a) The $36$-${\rm VBC}_{1}$ possesses only a single axis of reflection 
symmetry, which bisects the sides of the shaded hexagon; consequently, its 
symmetry group is ${\rm C}_{1v}$. It has $38$ classes of bonds. 
(b) The $36$-${\rm VBC}_{1}'$ has the same symmetry as $36$-${\rm VBC}_{1}$, 
but its reflection symmetry axis passes through a vertex of the shaded hexagon;
we shall denote its symmetry group as ${\rm C}_{1v}'$, to distinguish it from 
that of $36$-${\rm VBC}_{1}$. It has $36$ classes of bonds. Note that the $36$-${\rm VBC}_{2}$ (which has no symmetry at all) has not been drawn. 
Its symmetry group is just the identity {\it E}. Consequently, it has $72$ 
distinct classes of bonds.}
\label{fig:36-site-VBC-third-set}
\end{figure}

\subsection{$36$-site unit cell VBCs}
The building blocks of the kagom\'e lattice can take nontrivial forms such as 
a $2\sqrt3 \times 2\sqrt3$ expansion of the elementary $3$-site unit cell, 
thus giving rise to a tilted $36$-site unit cell. It was shown 
in~\cite{Marston-1991,Nikolic-2003} that such a construction maximizes the 
density of hexagons on which dimer resonances occur, thereby lowering 
the energy. This $36$-site unit cell VBC was studied numerically using series 
expansion~\cite{Singh-2007,Singh-2008} and MERA~\cite{Evenbly-2010}, which found it to 
be a good approximation to the ground state of the NN spin-$1/2$ QHAF.
Similar conclusions were also obtained 
from a QDM study~\cite{Poilblanc-2010,Schwandt-2010,Poilblanc-2011}. 
Motivated by these findings we classify all $36$-site unit cell VBC patterns on the 
kagom\'e lattice, which leads to the identification of nine symmetry distinct VBCs;
see figures~\ref{fig:36-site-VBC-first-set},~\ref{fig:36-site-VBC-second-set} 
and~\ref{fig:36-site-VBC-third-set} for their NN patterns. 

In our previous work~\cite{Iqbal-2010} we studied the HVBC state (see 
figure~\ref{fig:HVBC-36}) by using projected wave functions and found it to be 
higher in energy compared to the gapless SLs. However, the symmetry 
of the VBC identified in QDM studies~\cite{Poilblanc-2010,Schwandt-2010,Poilblanc-2011} 
is that of the ${\rm HVBC}_{0}$ state (see figure~\ref{fig:HVBC_0-36}), which 
has a lower symmetry compared to the HVBC state. In section~\ref{sec:numerics}, 
we study the possibility of a ground state realization of the ${\rm HVBC}_{0}$ state for
the NN and NNN spin-$1/2$ QHAF. 

\subsection{General remarks on the VBC classification}
It is worth mentioning that this VBC classification (for a given unit cell) 
is based on very general considerations of symmetry {\it only} and hence is 
{\it not} dependent on the formalism in which one studies these phases. 
In principle, it is possible to translate its construction from one language 
(e.g. QDM) into another (e.g. Schwinger fermions or bosons) for a VBC with a 
given symmetry. Moreover, within a given framework there can be different 
ways of constructing wave functions for a given VBC, consistent with its 
symmetry group. Firstly, one can add amplitudes beyond NN, consistent with 
the VBC symmetry group. Since we will study these phases within a slave 
particle approach, one can construct at the naive level simple mean-field wave
functions or go much beyond mean-field and include the effects of full 
projection. At a next level, it is possible to improve the wave function by 
applying the Hamiltonian operator on it a given number of times and 
considering an optimized linear superposition of these wave functions with the
original projected wave function. It is also worth noting that this hierarchical sorting of 
VBCs in each fixed symmetry sector also greatly eases the numerical search for a 
possible VBC stabilization as the ground state of the spin-$1/2$ QHAF.

\section{Numerical Results}
\label{sec:numerics}

We study the energetics of SL and VBC phases for the spin-$1/2$ QHAF using 
Gutzwiller projected fermionic wave functions with the variational quantum Monte 
Carlo technique. Our variational calculations are performed on clusters with $432$ 
(i.e. $3\times12\times 12$) or $576$ (i.e. $36\times 4\times 4$) sites
and mixed periodic-antiperiodic boundary conditions which ensure non-degenerate
mean-field wave functions at half filling. The large size of the cluster 
ensures that the spatial modulations induced in the observables by breaking of 
rotational symmetry (due to mixed boundary conditions) remain smaller than the
uncertainty in the Monte Carlo simulations.

Among the class of NN fully symmetric and gapless SLs, the U($1$) Dirac SL 
($[0,\pi]$ SL) has the lowest energy for the NN spin-$1/2$ QHAF. On a $432$-site cluster its energy per site is $E/J_{1}=-0.428~63(2)$ and the uniform RVB 
SL ($[0,0]$ SL) has a slightly higher energy per site, 
$E/J_{1}=-0.412~16(1)$~\cite{Ran-2007}. For the $576$-site cluster these values 
are $E/J_{1}=-0.428~66(1)$ for the U($1$) Dirac SL and $E/J_{1}=-0.411~97(1)$ 
for the uniform RVB SL~\cite{Iqbal-2010}. Upon inclusion of NNN hopping 
amplitudes, one gets the extended U($1$) Dirac SL or the extended uniform RVB 
SL, which are labeled by one additional flux through a plaquette of 
the type `$234$' in figure~\ref{fig:DSL-ansatz}, the flux through the other 
triangular plaquette formed by NNN bonds only is then fixed. Hence, the 
extended Dirac SL can be either the $[0,\pi;\pi,0]$ or the $[0,\pi;0,\pi]$ SL 
and analogously the extended uniform RVB SL can be either the $[0,0;\pi,\pi]$ 
or the $[0,0;0,0]$ SL~\cite{Iqbal-2010}.
For the NN spin-$1/2$ QHAF these extended SLs have a slightly lower energy, 
but they perform much better for the $J_{1}-J_{2}$ spin-$1/2$ QHAF, see figure~\ref{fig:VBC-J2}.

\begin{figure}
\centering
\subfloat[]{\label{fig:CVBC-op}\includegraphics[width=0.48\columnwidth]{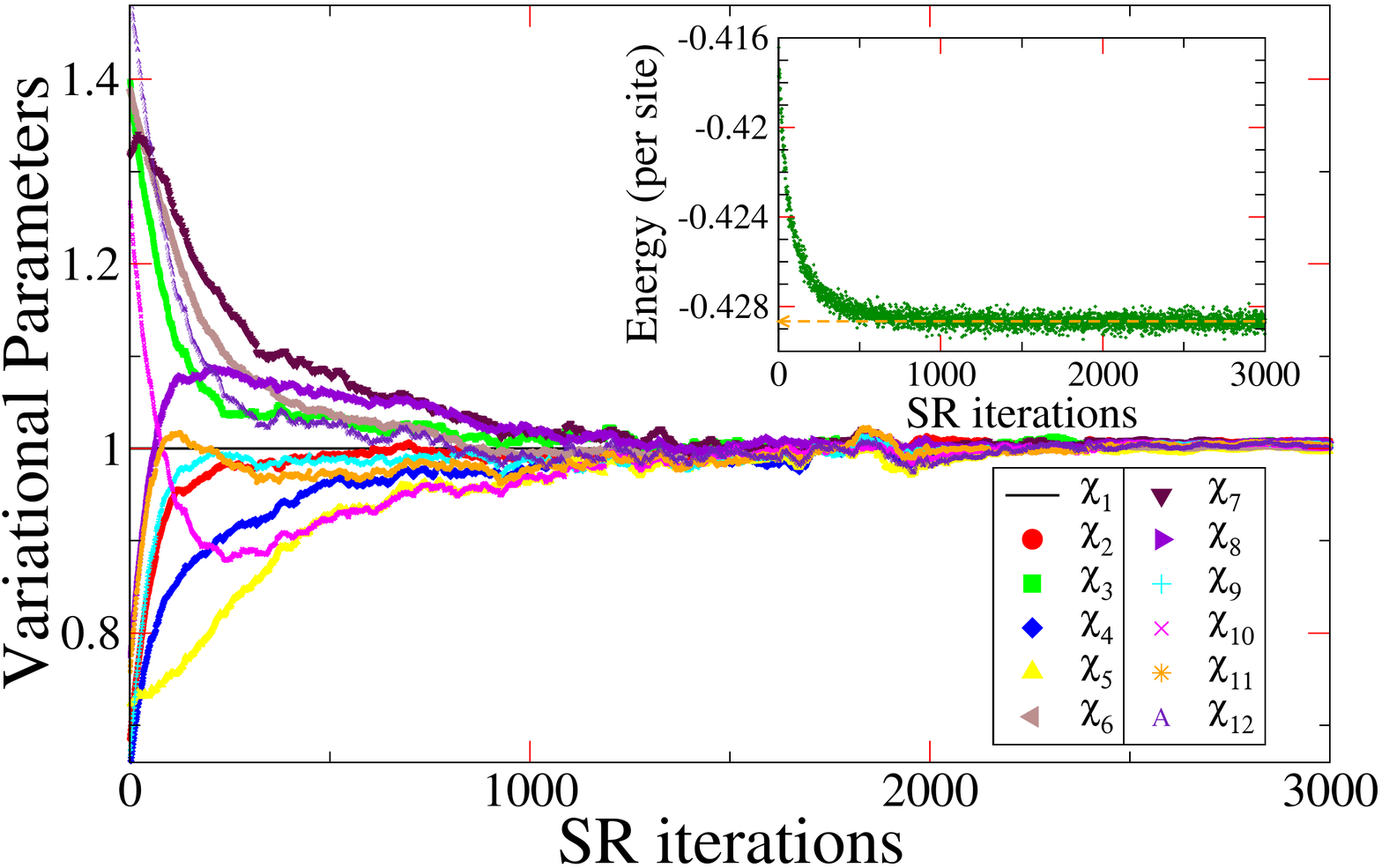}}\hspace{0.25cm} 
\subfloat[]{\label{fig:DVBC-op}\includegraphics[width=0.48\columnwidth]{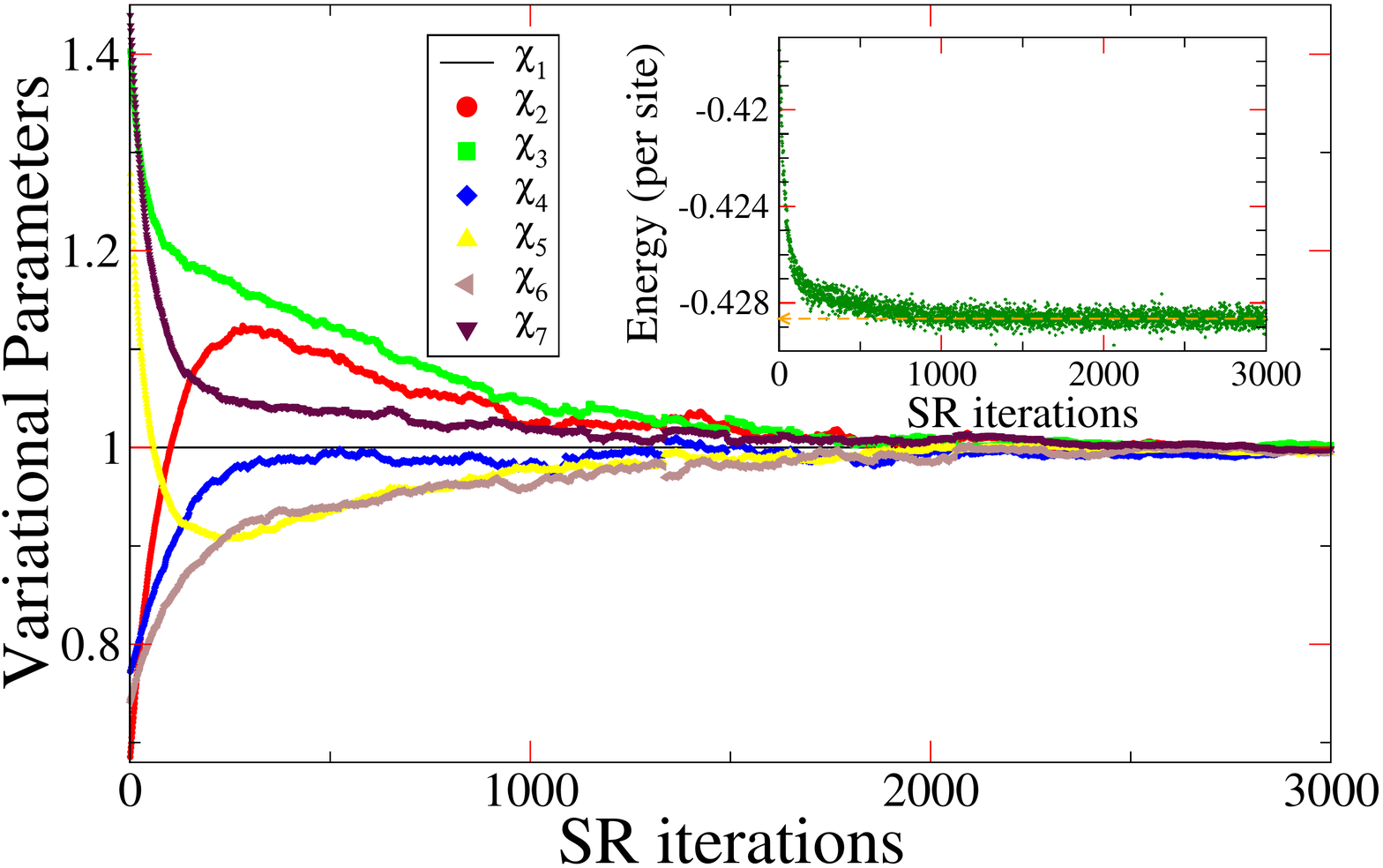}}\vspace{0.0cm}
\subfloat[]{\label{fig:VBC3-op}\includegraphics[width=0.48\columnwidth]{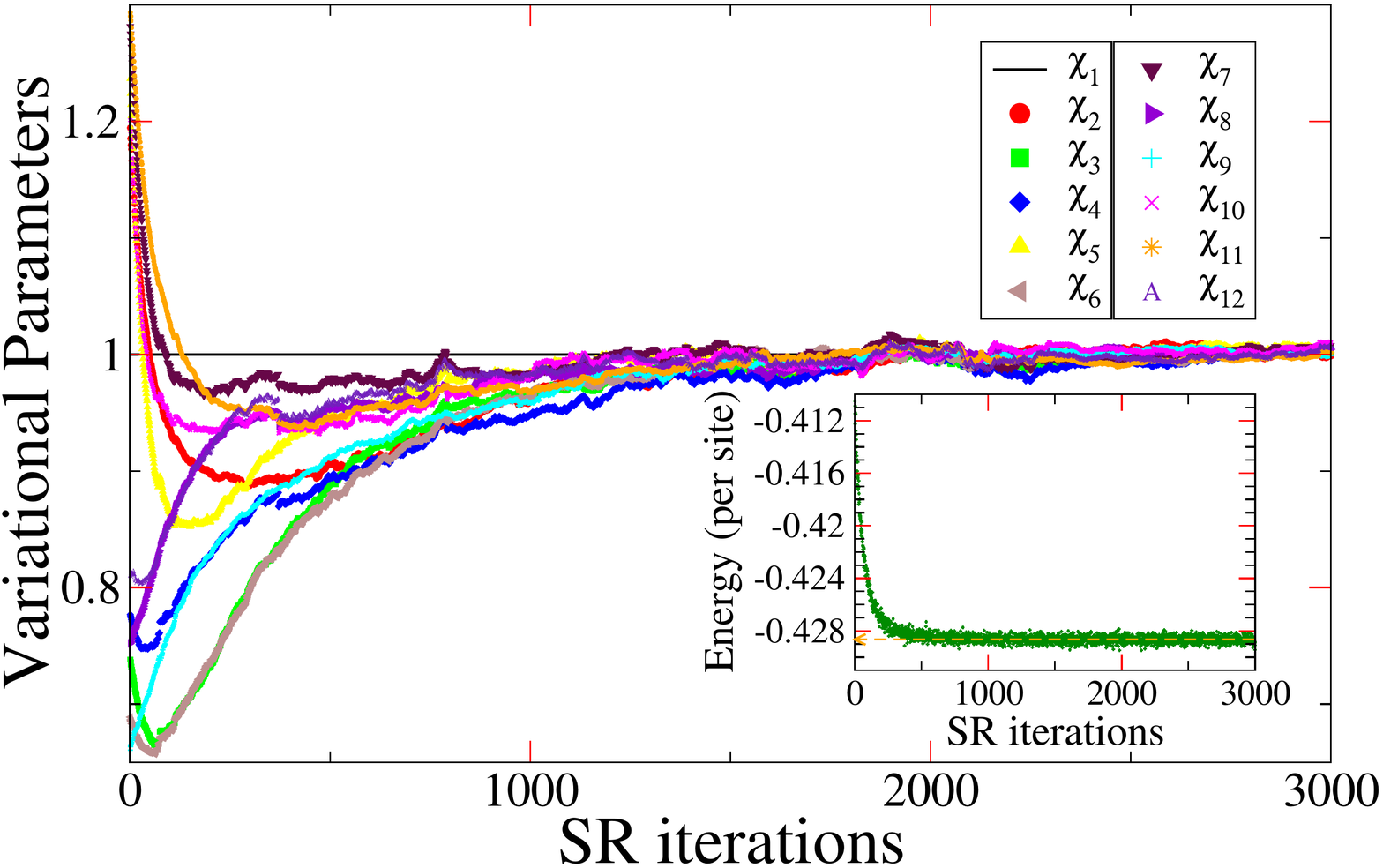}}\hspace{0.25cm}
\subfloat[]{\label{fig:HVBC0-op}\includegraphics[width=0.48\columnwidth]{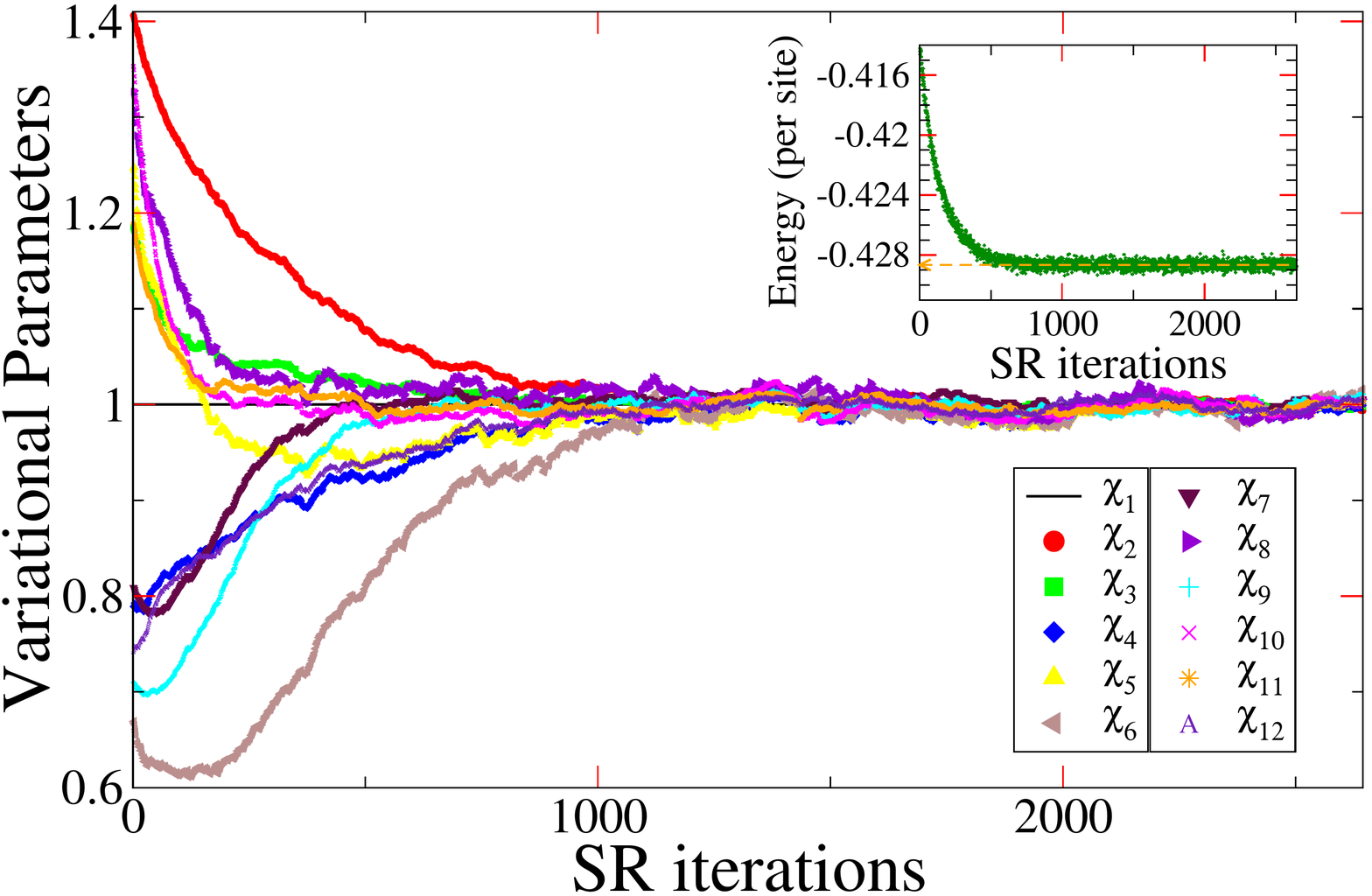}}
\caption{A typical variational Monte Carlo optimization run for the CVBC (a), 
DVBC (b), ${\rm VBC}_{3}$ (c) and ${\rm HVBC}_{0}$ (d) wave functions, for the
NN $S=1/2$ QHAF. The variational parameters $\chi_{\lambda}$ and energy 
(insets) are shown as a function of Monte Carlo iterations. The NN U($1$) Dirac
SL corresponds to $|\chi_{\lambda}|=1$. On starting from different sets of 
initialized parameter values we return back (within error bars) to the U($1$) 
SL. The optimized parameter values are obtained by averaging over a much 
larger number of converged Monte Carlo iterations than that shown above.}
\label{fig:op-curves}
\end{figure}

\subsection{Results on the stability of gapless SLs towards VBC perturbations}
We carried out an extensive numerical study of the local and global
stability of the NN U(1) Dirac and uniform RVB SL toward dimerizing into 
all $6$-, $12$- and $36$-site unit cell VBCs. In cases where we did find 
dimerization with NN bond amplitudes, we added second NN bond amplitudes to 
the SL and VBC ansatz (consistent with symmetries), since this led to a 
significant gain in energy. Our main focus was on the CVBC (figure~\ref{fig:CVBC-6}),
DVBC (figure~\ref{fig:DVBC-12}), ${\rm VBC}_{3}$ (figure~\ref{fig:VBC_3-12}) and 
${\rm HVBC}_{0}$ (figure~\ref{fig:HVBC_0-36}) states, since these have been 
identified as ground states of the spin-$1/2$ QHAF in other studies. 
We perform our analysis by first fixing a background flux corresponding to 
the SL liquid whose stability we wish to study. Then, we introduce an amplitude
modulation of $\chi_{ij}$ consistent with the symmetries of the VBC, i.e. 
bonds belonging to the same class (color/line marking in 
figures~\ref{fig:CVBC-6}, \ref{fig:12-site-VBC-first-set}, 
\ref{fig:12-site-VBC-second-set}, \ref{fig:36-site-VBC-first-set},~\ref{fig:36-site-VBC-second-set} 
and~\ref{fig:36-site-VBC-third-set}) have the same amplitude 
($\chi_{\lambda}$), which is set to different values for different classes. 
Starting from an arbitrary unbiased point ($\chi_{\lambda}$'s) in the 
variational space we perform an optimization of the wave function to obtain 
the lowest energy state~\cite{Sorella-2005,Yunoki-2006}.

\begin{figure}
\centering
\includegraphics[width=0.6\columnwidth]{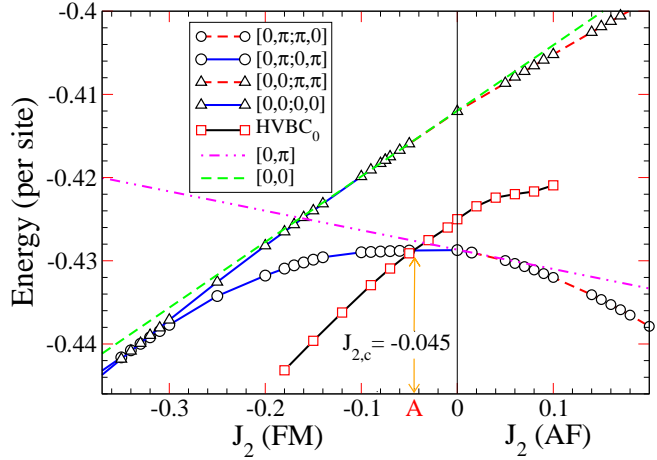}
\caption{Energy versus $J_{2}$ for SLs and the ${\rm HVBC}_{0}$ state 
(see figure~\ref{fig:HVBC_0-36}). The ${\rm HVBC}_{0}$ state becomes the 
lowest in energy for $J_{2}\lessapprox-0.045$. Error bars are smaller than the
symbol sizes.}
\label{fig:VBC-J2}
\end{figure}

\subsubsection{The case of the U($1$) Dirac SL}
\label{sec:Dirac-stability}
For the NN spin-$1/2$ QHAF, the variation of parameters and energy in the 
Monte Carlo optimization for the four competing VBCs (regarded as a 
dimerization of the U($1$) Dirac SL) mentioned above is given in 
figure~\ref{fig:op-curves}. As can be clearly seen, the energy converges 
neatly to the reference value of the U($1$) Dirac SL, and all the 
parameters converge to $\chi_{\lambda}=1$ (within error bars) after averaging 
over a sufficient number of converged Monte Carlo steps; thus the translation 
symmetry associated with the SL is restored. In fact, we performed these 
calculations for all $6$-, $12$- and $36$-site VBCs and found that in each
case the U($1$) Dirac SL is stable towards opening a gap and destabilizing 
into any of these VBCs. This remarkable stability (for all VBCs) 
is also preserved upon addition of a NNN ($J_{2}$) super-exchange coupling in
the Hamiltonian of both AF and FM type. We verified these results by doing many 
optimization runs starting from different initial values of the parameters in 
the respective variational spaces. Thus, we can safely conclude that the 
U($1$) Dirac SL has the lowest variational energy among all proposed competing
VBC states, at least within the Schwinger fermion representation of the spin 
model for $J_{2}$ greater than a certain critical value $J_{2,{\rm c}}$, 
which is given and discussed in the ensuing text. 

\subsubsection{The case of the uniform RVB spin liquid}
\label{sec:RVB-stability}
We now shift our focus to the uniform RVB SL and address the question of its 
stability. For the NN and NNN (AF and FM) spin-$1/2$ QHAF, we find that 
{\it all} $6$- and $12$-site unit cell VBCs have a higher energy compared to 
the uniform RVB SL. However, interestingly enough, for the NN spin-$1/2$ QHAF, 
this NN uniform RVB SL opens up a gap and destabilizes into a $36$-site unit 
cell VBC, namely the ${\rm HVBC}_{0}$ state (see figure~\ref{fig:HVBC_0-36}).
The gain in energy due to dimerization becomes more pronounced on addition 
of second NN hopping amplitudes to the wave function which are consistent with 
${\rm C}_{6}$ symmetry. On adding a NNN super-exchange coupling of FM type to 
the Hamiltonian and following this second NN ${\rm HVBC}_{0}$ state 
(now, a dimerization of the extended uniform RVB SL), one 
finds that it becomes the lowest in energy for $J_{2}\lessapprox-0.045$ 
(see point A in figure~\ref{fig:VBC-J2}), consistent with the findings 
in~\cite{Ralko-2010}. It is worth noting that the symmetry of this VBC is precisely that of the VBC 
identified in the QDM study~\cite{Poilblanc-2010,Schwandt-2010,Poilblanc-2011}
and has a lower symmetry compared to the HVBC state that was previously 
studied by us with similar conclusions~\cite{Iqbal-2010}. The flux pattern of this VBC consists 
of $0$ flux through all elementary triangles, hexagons and a $\pi$ flux through the `$234$ 
plaquettes (see figure~\ref{fig:DSL-ansatz}) inside the perfect hexagons only.
 The lower symmetry of the ${\rm HVBC}_{0}$ compared to the HVBC implies a larger variational 
space of hopping amplitudes and consequently a lower energy that is seen from the fact that the level crossing or the onset of VBC order is shifted
from $J_{2}\approx-0.09$~\cite{Iqbal-2010} for HVBC to $J_{2,{\rm c}}\approx-0.045$ 
for ${\rm HVBC}_{0}$ state. Thus our results still point to a gapless ground 
state for $J_{2}\gtrapprox-0.045$, which is along the lines of our previous 
work~\cite{Iqbal-2010,Iqbal-2011}.

\section{Conclusions and discussions}
In this paper, we enumerated all $6$-, $12$- and $36$-site unit cell VBCs 
based on symmetry considerations alone and subsequently investigated 
the possibility of stabilizing any of these VBCs in the NN and NNN spin-$1/2$ 
QHAF on a kagom\'e lattice. We found that the U($1$) Dirac SL is remarkably 
robust toward dimerizing into any of these VBCs, for both the NN and NNN 
spin-$1/2$ QHAF. However, the uniform RVB SL dimerizes into a $36$-site unit 
cell VBC, which becomes the lowest in energy on addition of a very weak FM 
coupling, $J_{2,{\rm c}}\approx-0.045$. Our systematic and thorough numerical 
investigation brings us to the conclusion that, at least within the Schwinger 
fermion approach to the spin model, the U($1$) Dirac SL has the best 
variational energy for $J_{2}\gtrapprox-0.045$. The conflict between our results, 
which point to a gapless ground state in this region, and those obtained by exact
diagonalizations and DMRG calculations, which instead suggested the presence 
of a fully gapped spectrum, remains open and deserves further investigation. 
One possible direction would be to include vison dynamics in the 
projected wave functions~\cite{Motrunich-2011}, which may be necessary to 
capture topological order faithfully. Another step would be to improve our variational
wave functions based on the application of a few Lanczos 
steps~\cite{Sorella-2001} and then perform an approximate fixed-node 
projection technique. The possibility that an unconventional
VBC breaking time-reversal symmetry is stabilized as the ground state cannot be ruled 
out~\cite{Poilblanc-2011}. Finally, we mention that VBC order might also 
set in via confinement transitions of the $\mathbb{Z}_{2}$ SLs~\cite{Huh-2011},
this remains to be investigated numerically.

\ack

YI and DP acknowledge support from the `Agence Nationale de la Recherche' under grant no.~ANR~2010~BLANC~0406-0.  We are grateful for the permission granted to access the HPC resources of CALMIP under the allocation 2012-P1231.

\end{document}